\documentclass[12pt]{article}
\usepackage{graphicx}
\usepackage{amssymb}
\usepackage{amscd}
\usepackage{amsmath}
\usepackage{epstopdf}
\usepackage{cite}

\begin{document}
\begin{titlepage}

{\hbox to\hsize{\hfill October 2018 }}

\bigskip \vspace{3\baselineskip}

\begin{center}
{\bf \large 
Electroweak monopoles and electroweak  baryogenesis \\ }

\bigskip

\bigskip

{\bf Suntharan Arunasalam, Daniel Collison and Archil Kobakhidze  \\ }

\smallskip

{ \small \it
ARC Centre of Excellence for Particle Physics at the Terascale, \\
School of Physics, The University of Sydney, NSW 2006, Australia \\
}

\bigskip
 
\bigskip

{\large \bf Abstract}

\end{center}
\noindent 
{\small We describe electroweak monopoles within the Born-Infeld extension of $SU(2)_L\times U(1)_Y$ electroweak theory. We argue for topological stability of these monopoles and computed their mass in terms of the Born-Infeld mass parameters. We then propose a new mechanism for electroweak baryogenesis which takes advantage of the following salient  features of the electroweak monopoles: (i) monopoles support extra CP violation in the topological sector of the electroweak theory; (ii) they mediate unsuppressed baryon number violating interactions; (iii) non-thermal production of monopoles during the electroweak phase transitions generates departure from thermal equilibrium.  We demonstrate that  the observed baryon asymmetry of the universe can be explained in our theory in the presence of electroweak monopoles of mass $M\sim 10^{4}$ TeV.}       

\bigskip
 
\bigskip

 \end{titlepage}

\section{Introduction}

In Ref. \cite{Arunasalam:2017eyu}, two of us have suggested that electroweak monopoles of mass $M\sim 10^{4}$ TeV, which are produced during the electroweak phase transition, may be responsible for the generation of the observed baryon asymmetry in the universe. The purpose of this paper is to study this scenario in more details within the Born-Infeld extension of the $SU(2)_L\times U(1)_Y$ electroweak theory, without introducing exotic new particles.

Previously it was thought that stable, regular (finite mass) monopoles do not exist in the electroweak theory. Some time ago, however, Cho and Maison proposed monopole (and dyon) solutions \cite{Cho:1996qd}, which are hybrids of the singular Dirac monopole \cite{Dirac:1948um} and the regular 'tHooft-Polyakov monopole \cite{tHooft:1974kcl, Polyakov:1974ek}. As we confirm below, these solutions are topologically stable. The serious drawback, however, is that the monopole static energy (mass) is divergent. In order to obtain finite mass monopoles, a short-scale modification of the electroweak theory is required. The mass of the monopole, then, is defined by the corresponding ultraviolet mass scale\footnote{Such extension of the electroweak theory can be achieved, for example, by embedding the $SU(2)_L\times U(1)_Y$ electroweak group into a larger (spontaneously broken) group. The prime example of this is the $SU(5)$ grand unified theory. $SU(5)$ monopoles, however, are too heavy to be phenomenologically viable.}. A relevant modification is provided by the fundamental string theory \cite{Fradkin:1985qd}, where gauge fields on D-branes are described by a non-linear Born-Infeld-type Lagrangian \cite{Born:1934gh}. Indeed, the Cho-Maison monopole acquires a finite mass in the Born-Infeld extension of the electroweak hypercharge gauge field \cite{Arunasalam:2017eyu}. In this paper we extend the previous study to the full electroweak theory.

There are a few interesting phenomena associated with the electroweak monopoles which have direct relevance for baryogenesis. First, the topologically non-trivial vacuum structure ($\theta$-vacuum) of the electroweak gauge theory implies that the electroweak monopoles also carry anomalous electric charge due to the Witten effect \cite{Witten:1979ey}. A new source of CP violation is thus introduced through these dyonic states. Second, monopole-antimonopole annihilations, while reducing the monopole abundance to an observationally  acceptable level, do also produce non-zero baryon ($B$) number due to the electroweak anomaly. Unlike instanton-induced $B$-violating processes, the monopole-mediated processes are not suppressed even at zero temperature \cite{Rubakov:1981rg, Callan:1982ah}. Finally, the monopoles are produced non-thermally during the electroweak phase transition via the Kibble mechanism \cite{Kibble:1976sj, Preskill:1979zi} and can drive the phase transition out of equilibrium \cite{Arunasalam:2017eyu}. In what follows we demonstrate that the presence of electroweak monopoles of mass $M\sim 10^{4}$ TeV sets the scene for the successful baryogenesis at the electroweak scale\footnote{For earlier attempts for baryogenesis with grand unified monopoles see \cite{Dixit:1991ym, Davis:1992ca}}.  

The paper is organised as follows. In the following section, we introduce the Born-Infeld extension of electroweak theory and discuss the electroweak monopole solutions. In Sec. 3, we discuss the $CP$ and $B+L$ violation induced by the electroweak monopoles.  In Sec. 4, we estimate baryon asymmetry generated by monopoles. We conclude in Sec. 5.

\section{The electroweak monopoles (dyons)}

Consider the Standard Model extended by Born-Infeld type kinetic terms for the hypercharge and non-Abelian $SU(2)_L$ gauge fields. The bosonic Lagrangian is given by 

\begin{multline}
\mathcal{L}=(D^\mu H)^\dagger D_\mu H - \frac{\lambda}{2}\Big(H^\dagger H- \frac{\mu^2}{\lambda} \Big)^2 +
\beta_1^2 \Bigg[1-\sqrt{1+\frac{1}{2 \beta_1^2} G_{\mu \nu}G^{\mu \nu}-\frac{1}{16 \beta_1^4}(G_{\mu \nu} \widetilde{G}^{\mu \nu})^2} \Bigg] 
\\ \\+\beta_2^2 \Bigg[1-\sqrt{1+\frac{1}{2 \beta_2^2} F_{\mu \nu}^a F^{a \mu \nu}-\frac{1}{16 \beta_2^4}(F_{\mu \nu}^a \widetilde{F}^{a \mu \nu})^2} \Bigg]
\end{multline}
where $D_\mu  \equiv (\partial_\mu -i\frac{g_2}{2}\tau^aA_\mu^a-i\frac{g_1}{2}B_\mu)$  $a=1,2,3$ is the SU$(2)_L$ $\times$ U$(1)_Y$ gauge covariant derivative, $H$ the electroweak doublet Higgs field and $A_\mu^a$ and $B_\mu$ the SU$(2)_L$ and U$(1)_Y$ gauge fields respectively. $F_{\mu \nu}^a$ are the SU$(2)_L$ field strength tensors and $B_{\mu \nu}$ the U$(1)_Y$  field strength tensor. For this analysis, the non-Abelian Born-Infeld term is defined by taking the trace of the field strength tensors under the square root.  The Born-Infeld parameters $\beta_1$ and $\beta_2$ with dimensions (mass)$^2$ control the non-linearity of the hypercharge and non-Abelian gauge fields respectively with $\beta_1 \rightarrow \infty$, $\beta_2 \rightarrow \infty$ recovering the Standard Model theory. 

The equations of motion for the fields read

\begin{equation}\label{fish1}
D_\mu (D^\mu H)=-\lambda\Big(H^\dagger H-\frac{\mu^2}{\lambda} \Big)H,
\end{equation}
\begin{equation} \label{fish2}
(\partial_\mu-i\frac{g_2}{2}\tau^aA_\mu^a)\Bigg[ \frac{F^{i \mu \nu}-\frac{1}{4\beta_2^2} (F_{\rho \sigma}^j \widetilde{F}^{j \rho \sigma})\widetilde{F}^{i \mu \nu}}{\sqrt{1+\frac{1}{2 \beta_2^2} F_{\rho \sigma}^jF^{j \rho \sigma}-\frac{1}{16 \beta_2^4}(F_{\rho \sigma}^j \widetilde{F}^{j \rho \sigma})^2} }\Bigg]=i\frac{g_2}{2}[H^\dagger \tau^i (D^\nu H)-(D^\nu H)^\dagger \tau^i H],
\end{equation}

\begin{equation} \label{fish3}
\partial_\mu \Bigg[ \frac{G^{ \mu \nu}-\frac{1}{4\beta_1^2} (G_{\rho \sigma} \widetilde{G}^{ \rho \sigma})\widetilde{G}^{ \mu \nu}}{\sqrt{1+\frac{1}{2 \beta_1^2} G_{\rho \sigma}G^{ \rho \sigma}-\frac{1}{16 \beta_1^4}(G_{\rho \sigma}\widetilde{G}^{\rho \sigma})^2} }\Bigg]=i\frac{g_1}{2}[H^\dagger (D^\nu H)-(D^\nu H)^\dagger H].
\end{equation}

\begin{figure}[t]
\centering
\includegraphics[width=13.5cm]{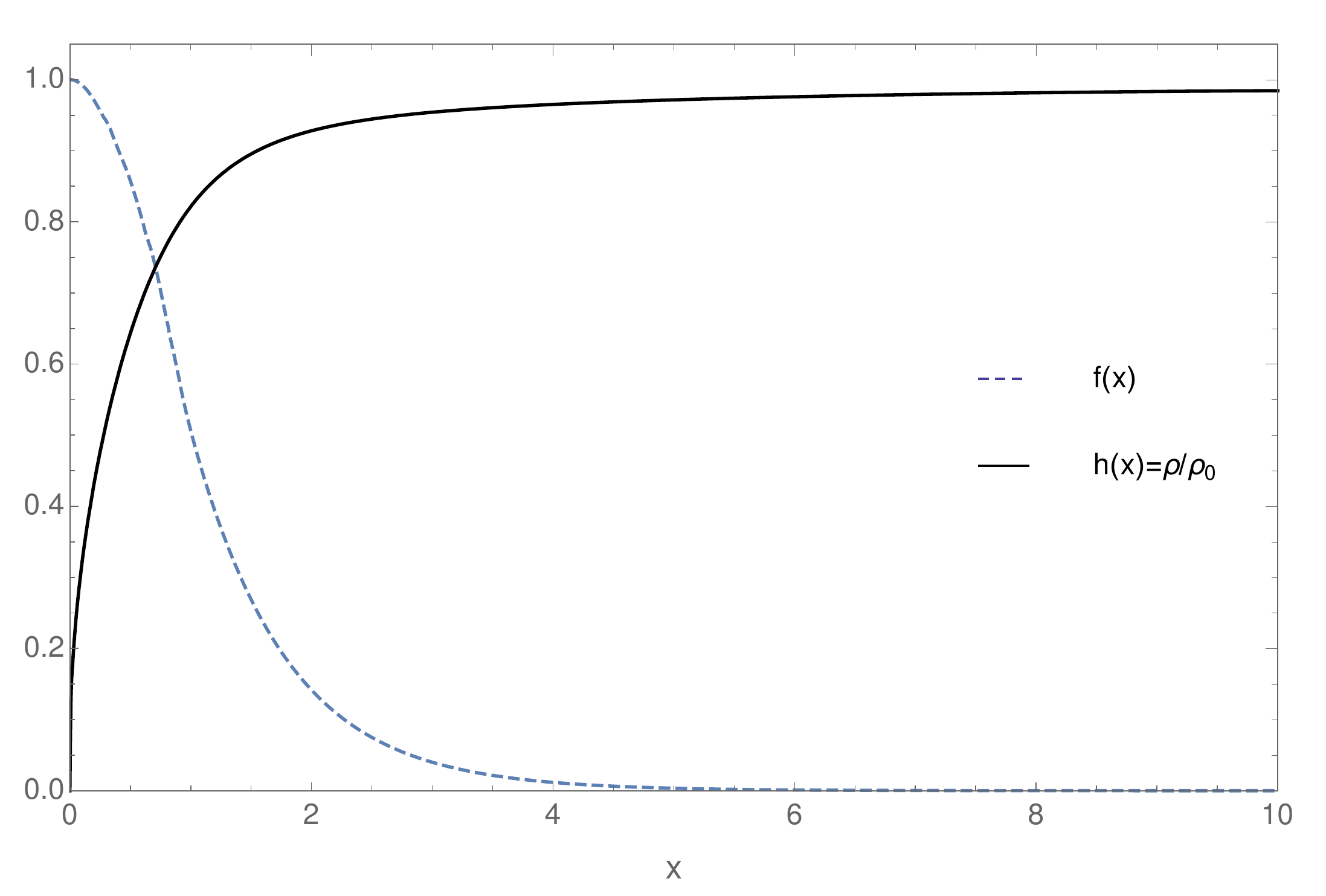}
\caption{\label{fishy} Solution to differential equations (\ref{diffone}) and (\ref{difftwo}) for $f$ and $\rho$. The equations where rewritten in terms of the dimensionless variable $x\equiv \mu r$ and $\rho$ was expressed as the dimesionless function $h \equiv \rho/\rho_0$. The dimensionless parameter $\alpha \equiv (g_2 \beta_2)/\mu^2 $ is fixed for this solution at $\alpha=10$ which corresponds to $\beta_2= (250 \hspace{0.2cm} \textnormal{GeV})^2$.}
\label{figure:graph}
\end{figure}

Consider the following ansatz

\begin{align}
H &=\frac{1}{\sqrt{2}}\rho(r) \zeta, \hspace{0.4cm} \zeta=i\begin{pmatrix}
   \sin(\theta/2)e^{-i\phi}  \\
  - \cos(\theta/2)
 \end{pmatrix}, \label{Hansatz} \\ \textbf{A}_\mu&=-\frac{1}{g_2} A(r)\partial_\mu t \hat{r}  +    \frac{1}{g_2}(f(r)-1)\hat{r}\times \partial _\mu \hat{r} ,
 \label{Aansatz}
\\
B_\mu&=-\frac{1}{g_1} B(r) \partial_\mu t -\frac{1}{g_1}(1-\cos \theta )\partial _\mu \phi. \label{Bansatz}
\end{align}
Here $\rho(r)$, $f(r)$, $A(r)$ and $B(r)$ are arbitrary functions of the radial coordinate $r$.  The functions $A(r)$ and $B(r)$ present the dyon solutions of this model. Pure magnetic monopole solutions are found by setting $A(r)=B(r)=0$.  In this case, Eq. (\ref{fish3}) is trivially satisfied and Eqs. (\ref{fish1}) and (\ref{fish2}) give coupled ODEs

\begin{equation}\label{diffone}
\rho''+\frac{2}{r} \rho'-\frac{f^2}{2r^2}\rho=\lambda\Big(\frac{\rho^2}{2}-\frac{\mu^2}{\lambda}\Big)\rho
\end{equation}
 \begin{equation}\label{difftwo}
f'' +f'\Big(\frac{2}{r}-\frac{R'}{R}\Big)-\frac{f^2-1}{r^2}f=\frac{g^2 R}{4r^2}\rho^2 f
\end{equation}
where 
 \begin{equation*}
R \equiv r^2 \sqrt{1+\frac{(f^2-1)^2}{g_2^2 \beta_2^2 r^4}+\frac{2{f'}^2}{g_2^2\beta_2^2 r^2}}.
\end{equation*}
The boundary values can be chosen as follows
\begin{align}
f(0) &= 1  &  \rho(0)  &= 0 \nonumber \\ f(\infty) &= 0      &   \rho(\infty) &= \rho_0=\sqrt{\frac{2\mu^2}{\lambda}}.
\end{align}
These equations can be solved using numerical methods, see Figure \ref{figure:graph}. 

The energy of this monopole solution is given by 
\begin{align}\label{fishyeleven}
    E &=4 \pi  \int_0^\infty{dr \beta_1^2 \bigg(\sqrt{r^4 +\frac{1}{g_1^2 \beta_1^2}} -r^2 \bigg)  +\beta_2^2 \bigg(\sqrt{r^4 +\frac{(f^2-1)^2}{g_2^2 \beta_2^2}+\frac{2f'^2 r^2}{g_2^2 \beta_2^2}} -r^2 \bigg)} \nonumber \\
      &\qquad + \frac{1}{2}(r \rho')^2+\frac{\lambda r^2}{8}(\rho^2-\rho_0^2)^2 +\frac{1}{4}f^2\rho^2.
  \end{align} 
The first term involving the Abelian Born-Infeld contribution may be expressed in closed form using elliptical integrals and is ultimately responsible for the energy being finite \cite{Arunasalam:2017eyu, Kim:2000}. It gives a contribution to the energy of $\approx 77.1 \sqrt{\beta_1} $. 
In the limit $\beta_2 \rightarrow \infty$ the rest of the energy contribution gives $\approx 2.8$ TeV.

Recent analysis of light by light scattering experiments at the LHC requires that the Born-Infeld parameter present in the usual Born-Infeld extension of QED be constrained by $\sqrt{\beta} \gtrsim 100$ GeV \cite{Ellis: 2017}.  Rewriting this constraint for the SU$(2)_L$ $\times$ U$(1)_Y$ electroweak Born-Infeld extension gives 

 \begin{equation}\label{fish11}
\frac{\sqrt{\beta_2}}{\sqrt[4]{\sin^4\theta_W+\cos^4\theta_W \big(\frac{\beta_2}{\beta_1}\big)^2}}  \gtrsim 100 \hspace{0.2cm}\textnormal{GeV} 
\end{equation}
by expanding the Born-Infeld terms in the Lagrangian in terms of the physical fields using the relations $B_\mu=\cos\theta_W A_\mu^{(em)}-\sin\theta_W Z_\mu$ and $A_\mu^3=\sin\theta_W A_\mu^{(em)} +\cos\theta_W Z_\mu$. 

In the limit  $\beta_2 \gg \beta_1$, $\sqrt{\beta_1}   \gtrsim  (100 \cdot \cos\theta_W)$ GeV $\approx 88$ GeV. In the opposite limit $\beta_2$ $\ll$ $\beta_1$, $\sqrt{\beta_2} \gtrsim  (100 \cdot \sin\theta_W)$ GeV $\approx 47$ GeV. Thus a lower bound on the Abelian U$(1)_Y$ contribution is $(77.1 \cdot 88)$ GeV $\approx 6.8$ TeV. In order to explore lower bounds for the monopole mass it is sensible to hold $\sqrt\beta_1 \approx 88$ GeV and vary $\beta_2$ which is unconstrained by the above analysis in the limit $\beta_2 \gg \beta_1$.

It is also possible to constrain the values for $\beta_2$ by considering contributions to vector boson scattering, in particular longitudinal $WW$ scattering. In the standard theory the leading order energy contribution of the basic single vertex scattering diagram does not contribute to the total amplitude. This is due to delicate cancellation with the higher order diagrams involving interactions with the Higgs, the photon and $Z$ bosons \cite{Fabbrichesi:2016, Endo:2016}. Without this cancellation, perturbative unitarity is spoiled in the Standard Model theory. With the Born-Infeld extension of SU$(2)_L$ this type of cancellation is not possible due the presence of terms such as $\frac{1}{\beta_2^2}(F^a_{\mu \nu} F^{a {\mu \nu}})^2$ in the expansion of the square root which also, to lowest order, contains the basic $WW$ scattering vertex in the form of a product of four $\partial_\mu W$ terms. Naively the amplitude should scale as $A^{BI}_{WW\rightarrow WW} \propto  E_W^8 /({\beta_2^2 m_W^4})$ where $E_W$ is the center-of-mass energy of the $W$ and $m_W$ its mass.  

To ensure perturbative unitarity is respected

\begin{equation}
E_W^8 /({\beta_2^2 m_W^4})<1.
\end{equation}
At centre-of-mass energies $ E \sim \mathcal{O}$ (TeV) this corresponds to a constraint on $\beta_2$


\begin{equation}\label{fishy70}
\sqrt{\beta_2} \gtrsim 10^5 \hspace{0.1cm} \textnormal{GeV}.
\end{equation}

Together with the lower bound on the U$(1)_Y$ contribution the resulting lower bound on the total monopole mass is $\sim 9-11$ TeV.

\subsection{The topological stability of the monopole}

In this section we discuss  topology of the monopole solution considered in the previous section. This topic still raises a controversy, with continuing claims in the literature on non-existence of topologically stable monopoles in the standard model. The argument against electroweak monopoles is based on the following consideration. The field configurations away from the monopole core approach their vacuum configurations. In particular, the Higgs fields vacuum configurations satisfy, 
\begin{equation}\label{st1}
H^{\dagger}H\equiv \phi_1^2+\phi_2^2+\phi_3^2+\phi_4^2\stackrel{r\to\infty}{=}\rho_0^2~,
\end{equation}
and hence the Higgs vacuum manifold is considered a 3-sphere, $S^3$, provided that the real components $\phi_i$ of the electroweak doublet are regular functions.  If so, the Higgs configuration for the monopole solution represents a map from the $S^2$ boundary of the 3-dimensional space into an $S^3$ vacuum manifold. This map is indeed homotopically trivial, $\pi_2(S^3)=1$, therefore, electroweak monopoles carry no topological charge. However, the Higgs field configuration of Eq. (\ref{Hansatz}) as well as hypercharge field Eq. (\ref{Bansatz}) have string singularity along the negative z-axis, i.e. at $\theta=\pi$. Therefore, the topology of the monopole solution requires more careful consideration. 

To begin, we note that the components of the electroweak Higgs doublet $z_1\equiv \phi_1+i\phi_2$, $z_2\equiv \phi_3+i\phi_4$ with $(z_1,z_2)^{T}\neq (0,0)^{T}$, are defined up to $U(1)_Y$ gauge transformations, i.e. $(z_1,z_2)^{T}\equiv (\lambda z_1, \lambda z_2)^{T}$, $\lambda \in U(1)_Y$, and hence can be viewed as a coordinates on the complex plane $\mathbb{C}^2$. Next, following the Wu-Yang construction{\cite{Wu:1975es}, we remove the string singularity by using the gauge freedom and defining the monopole solution on two different patches as follows: 
\begin{eqnarray}
H_N=i\frac{\rho(r)}{\sqrt{2}}\begin{pmatrix}
   \sin(\theta/2)e^{-i\phi}  \\
  - \cos(\theta/2)
 \end{pmatrix}~,~~
B^N_\phi=-\frac{1}{g'}\frac{1-\cos \theta }{r\sin\theta}
\label{st2} 
\end{eqnarray}
for $0\leq\theta\leq \pi/2$, and    
\begin{eqnarray}
H_S=i\frac{\rho(r)}{\sqrt{2}}\begin{pmatrix}
   \sin(\theta/2)  \\
   -\cos(\theta/2)e^{i\phi}
 \end{pmatrix}~,~~
B^S_\phi =\frac{1}{g'}\frac{1+\cos \theta }{r\sin\theta}
\label{st3} 
\end{eqnarray}
for $\pi/2\leq \theta \leq \pi$. We are free to define new coordinates on two patches of the complex plane $\mathbb{C}$: $\zeta_N=(z^N_1/z^N_2, 1)^{T}=(-\tan{(\theta/2)}e^{-i\phi}, 1)$ and $\zeta_S=(1, z^S_2/z^S_1)^{T}=(1, -\cot{(\theta/2)}e^{i\phi})^{T}$ as $z_2^N, z_1^S\neq 0$. The coordinate transition between these two charts at the equatorial plane $\theta =\pi/2$ is evidently given by the holomorphic function, $e^{i\phi}$, and, therefore, these two charts actually cover complex projective line $\mathbb{CP}^1$ (Riemann sphere) by definition, rather than just $\mathbb{C}^2$. Hence the Higgs monopole configuration represents a mapping which is topologically non-trivial, $\pi_2(\mathbb{CP}^1)=\pi_2(S^2)=\mathbb{Z}$.  

\section{Violation of $CP$ and $B+L$ by the electroweak monopoles}
Although we describe the electroweak monopole as a classical solution, a number of important properties follow from  quantum considerations. As it is well-known, despite the existence of (restricted)  $SU(2)_L$ electroweak instantons in the vacuum sector of the electroweak theory and the associated topological classification of degenerate gauge vacua, the transition between those vacua are forbidden due to the chiral nature of the electroweak theory. This means, CP-violating electroweak $\theta$-terms, 
\begin{equation}
{\cal L}_{\theta}=\theta_2F_{\mu\nu}^a\tilde F^{a\mu\nu}+\theta_1B_{\mu\nu}\tilde B^{\mu\nu}~,
\label{cp1}
\end{equation}
are actually unobservable: the $\theta_1$-term is a `true' total derivative due to the trivial topology of $U(1)_Y$ gauge vacuum and the $\theta_2$-term can be removed by $B+L$ rotation of quarks and leptons. In contrast, both of these terms are supported by electroweak monopoles, which due to the Witten effect \cite{Witten:1979ey} acquire an electric charge proportional to $\theta_{em}=\theta_1\cos^2\theta_W+\theta_2\sin^2\theta_W$ and become dyons\footnote{A constraint on the $\theta_{em}$ may arise from the contribution of virtual dyons to electric dipole moments of known particles (see, e.g. Ref.  \cite{Murray:1998bu}). We verify, however, that current experiments are not sensitive to the $CP$ violation induced by the heavy electroweak monopoles, discussed in this paper.}. Importantly, due to the anomalous violation of $B+L$, we can remove one of the phases only, the another one would contribute to physical processes. We rotate quarks and leptons such that Eq. (\ref{cp1}) becomes:  
\begin{equation}
{\cal L}_{\theta}=\theta_{ew}F_{\mu\nu}^a\tilde F^{a\mu\nu}~,
\label{cp2}
\end{equation}
where $\theta_{ew}=\theta_2-\theta_1$ is non-zero, unless $\theta_1$ and $\theta_2$ conspire to cancel each other. Hence, together with the existence of monopoles additional $CP$ violation is inevitable in the standard model\footnote{An additional $CP$ violation in the standard model may also be induced by gravitational instantons \cite{Deser:1980kc, Arunasalam:2018eaz}. }.   

The electroweak gauge fields interpolate between topologically inequivalent vacuum configurations related by large gauge transformations $g\in SU(2)_L$ giving rise to the $\theta_{ew}$-vacuum structure. Denoting by $g$ a transformation that gives a unit topological charge of the electroweak instanton, an anomalous processes with $\Delta B=\Delta L=3$ are also induced. These processes are exponentially suppressed, and can be relevant only in the cosmological setting at high temperatures. 
In the monopole sector, the corresponding gauge transformation of the perturbative monopole state $\vert M,0 \rangle$, i.e., $U[g]\vert M,0 \rangle $ would be a state which carries $B=L=3$. The physical monopole state than is a linear superposition of the form \cite{Pak:1980sq} (see also the second reference in \cite{Rubakov:1981rg}):
 \begin{equation}
\vert M,\theta_{ew} \rangle=\sum_{n=-\infty}^{n=+\infty} e^{in\theta_{ew}}(U[g])^n \vert M,0 \rangle~. 
\label{cp3}
\end{equation}

Anomalous violation of $B+L$ in scatterings involving monopoles or monopole-antimonopole annihilation processes can be easily understood in this language. E.g., monopole-antimopole pair that carries $\Delta n=1$ topological charge, would annihilate into 9 quarks and 3 leptons, giving rise to $\Delta B=\Delta L=3$. As shown by Callan \cite{Callan:1982ah} and Rubakov \cite{Rubakov:1981rg} a long time ago, the monopole mediated baryon number violating processes are not suppressed even at zero temperature. Furthermore, due to the $CP$ violation by $\theta_{ew}$, $(B+L)$-violating processes are $CP$ non-invariant. This can be captured through the $\theta_{ew}$-dependence of the Hamiltonian in the anomalous commutator with $Q_{B+L}$ charge operator:
  \begin{eqnarray}
\dot Q_{B+L}&=&i[Q_{B+L}, H(\theta_{ew})] \nonumber \\
&=&i[Q_{B+L}, H(\theta_{0})]+i\theta_{ew}\left[Q_{B+L}, \left.\frac{\partial H}{\partial \theta_{ew}}\right \vert_{\theta_{ew}=0}\right]+{\cal O}(\theta_{ew}^2). 
\label{cp4}
\end{eqnarray}   
The first commutator on the second line of the above equation describes the usual $CP$-conserving $B+L$-violation due to the anomaly, while the second term is $CP$-odd and, hence, differs by sign for particles and antiparticles. In the next section, we incorporate these findings to describe cosmological generation of matter-antimatter asymmetry during the electroweak phase transition accompanied by production and subsequent annihilation of electroweak monopoles.

\section{Generation of baryon asymmetry by the electroweak monopoles }

In \cite{Arunasalam:2017eyu}, two of us discussed the effects of these monopoles on the electroweak phase transition. During the phase transition, these monopoles will be copiously produced through the Kibble Mechanism \cite{Kibble:1976sj}. This subsequently adds to the Gibbs free energy of the broken phase, resulting in a stronger phase transition. It was shown in \cite{Arunasalam:2017eyu} that estimating the initial density of these monopoles as $d_c^{-3}$, where $d_c$ is the Coulomb capture distance, one obtains a strongly first order phase transition for $M\gtrsim 0.9\times 10^4$ TeV. In particular, this implies that the monopole production leads to the suppression of sphaleron processes, thus preventing the washout of the asymmetry. These monopoles and anti-monopoles subsequently annihilate with each other rapidly until they freeze out. It was shown that satifying BBN constraints requires that the monopole mass satisfies $M\lesssim 2.3\times 10^4$ TeV. 
We now compute the baryon asymmetry generated by the annihilation of these monopole-antimonopole pairs. This follows closely the calculations performed in \cite{Davis:1992ca} in the context of monopoles in the Langacker-Pi scenario. 

As discussed in the previous section, the rate of production of matter-antimatter asymmetry is proportional to $\theta$ and is given by:
\begin{align}
\frac{d \bar{n}_{B}}{dt}=-\kappa\theta\frac{dn_{M}}{dt}
\end{align}
where $\bar{n}_{B}$ is the difference in the number densities of matter and antimatter and $n_{M}$ is the number density of the monopoles. $\kappa$ is a parameter that describes the asymmetry generated in each collision. For example, $\Delta n=1$ would imply 9 quarks and 3 leptons are produced. As discussed above, the initial number density of the monopoles is given by
\begin{align}
n_{0}=\frac{1}{d_c^3}=\alpha_{EM}^3 T_c^3.
\end{align}
As computed in \cite{Preskill:1979zi}, following the rapid annihilation of these monopoles, they freeze out at a temperature, $T_f$, with a density of  
\begin{align}
n_f=\frac{\alpha_{EM}^3}{4\pi B} \frac{M}{CM_{P}}T_f^3
\end{align}
where $B=\left(\frac{3}{4\pi^2}\right)\zeta(3)\sum_i\left(\frac{hq_i}{4\pi}\right)^2\approx 3.5$ and $C=\sqrt{\frac{45}{4\pi^3 g_{\star}}}$ with $g_\star\approx 100$.	As can be seen, almost all the monopoles and antimonopoles annihilate with each other by the time they freeze out. Hence, we can estimate the asymmetry density generated to be:
\begin{align}
\bar{n}_{B}\approx\kappa\theta n_0=\kappa\theta\alpha_{EM}^3 T_c^3
\end{align}
In order to compute the asymmetry parameter, $\eta_{B}$, one must also consider the entropy density of the universe. The annihilation of the monopoles will resulting in reheating the universe which can be estimated as: $\rho_\text{rad}\to (1+\omega)\rho_\text{rad}$ where \begin{align}
\omega=\frac{\rho_m}{\rho_\text{rad}}=\frac{30 m\alpha_\text{EM}^3 T_c^3}{g_\star \pi^2 T_f^4}
\end{align}
For the monopoles satisfying the mass constraints discussed above, $0.03<\omega<0.07$.
Hence, as $s\to (1+\omega)^{\frac{3}{4}}s$, we see that  reheating has very little effect on the asymmetry parameter. Hence, computing the asymmetry parameter, one obtains:
\begin{align}
\eta_{B}=\kappa \theta \frac{n_0}{s}=\kappa\theta \frac{45\alpha_\text{EM}^3 T_c^3}{2\pi^2 g_\star T_f^3}
\end{align}
Hence, for $0.9\times 10^4 \text{ TeV}<M<2.3\times 10^4 \text{ TeV}$, we obtain $1.6\times 10^{-8}\kappa\theta
\leq\eta_{B}\leq 2.5\times 10^{-7} \kappa \theta$. Hence, empirical values for the asymmetry parameter $\eta_B\approx 10^{-10}$ can be accommodated for with $\kappa\theta_{ew}\sim 10^{-3}-10^{-2}$. 

\section{Conclusion and discussion}

The electroweak monopoles discussed in this paper are necessarily present in the Born-Infeld extension of the standard model and invoke many remarkable properties. In particular, they support new $CP$ violating parameter $\theta_{ew}$ and thus contribute to the electric dipole moments of known particles. In addition, they induce baryon and lepton number violating processes, which unlike the electroweak instanton mediated similar processes are not suppressed even at zero temperatures. Taking all the relevant experimental constraints into account, we have estimated the lowest mass of such electroweak monopole to be $\sim 9-11$ TeV. 

The electroweak monopoles discussed in this paper differ in various aspects with the Nambu solution of  eletroweak Z-strings \cite{Nambu:1977ag, Achucarro:1999it}, which connect a pair of monopole and antimonopole. These solutions are not stable, however, and in fact may decay to quarks and leptons with non-zero $B+L$ number. Unlike the electroweak monopoles, the electroweak Z-strings do not support the electroweak $\theta_{ew}$-term \cite{Vachaspati:1994xe}.   

Heavy electroweak monopoles, $M\sim 10^4$ TeV, may play significant role in the early universe cosmology. They are produced non-thermally during the electroweak phase transition and drive it to be a sufficiently strongly first-order transition. Monopole-antimonopole annihilations not only reduce the monopole abundance down to an acceptable level, but also produce sufficient matter-antimatter asymmetry in the universe. 

While the lightest electroweak monopoles can be directly produced at future high energy colliders (e.g., a 100 TeV collider), the cosmologically interesting heavy monopoles will not be accessible for any feasible future machine. Nevertheless, they can be searched in various astrophysical experiments. We plan more detailed study of collider phenomenology and astrophysical manifestations of electroweak monopoles elsewhere.      

 \paragraph{Acknowledgements.} The authors would like to thank Zurab Berezhiani and Fabrizio Nesti for useful discussions. The work was partially supported by the Australian Research Council. AK also acknowledges the partial support from the INFN and the Simons Foundation (grant 341344, LA) during his visit at the Galileo Galilei Institute for Theoretical Physics, where part of this work has been carried out.

\end{document}